\documentclass[journal=jpclcd,manuscript=article]{achemso}

\usepackage{chemformula} 
\usepackage[T1]{fontenc} 

\author{Yichen Jin}
\affiliation{Department of Physics, Shanghai University, 99 Shangda Road, 200444 Shanghai, P. R. China}
\author{Mouhui Yan}
\affiliation{Department of Physics, Shanghai University, 99 Shangda Road, 200444 Shanghai, P. R. China}
\author{Tomislav Kremer}
\affiliation{Institut f\"ur Chemie und Biochemie, Freie Universit\"at Berlin, 14195 Berlin, Germany}
\author{Elena Voloshina}
\affiliation{Centre of Excellence ENSEMBLE3 Sp.\,z o.\,o., Wolczynska Str.\,133, 01-919 Warsaw, Poland}
\alsoaffiliation{Department of Physics, Shanghai University, 99 Shangda Road, 200444 Shanghai, P. R. China}
\alsoaffiliation{Institut f\"ur Chemie und Biochemie, Freie Universit\"at Berlin, 14195 Berlin, Germany}
\email{elena.voloshina@icloud.com}
\author{Yuriy Dedkov}
\affiliation{Centre of Excellence ENSEMBLE3 Sp.\,z o.\,o., Wolczynska Str.\,133, 01-919 Warsaw, Poland}
\alsoaffiliation{Department of Physics, Shanghai University, 99 Shangda Road, 200444 Shanghai, P. R. China}
\alsoaffiliation{Institut f\"ur Chemie und Biochemie, Freie Universit\"at Berlin, 14195 Berlin, Germany}
\email{yuriy.dedkov@icloud.com}

\title[]{Mott-Hubbard Insulating State for the Layered van der Waals FePX$_3$ (X:S, Se) As Revealed by NEXAFS and Resonant Photoelectron Spectroscopy}

\begin{document}








\begin{abstract}
A broad family of the nowadays studied low-dimensional systems, including 2D materials, demonstrate many fascinating properties, which however depend on the atomic composition as well as on the system dimensionality. Therefore, the studies of the electronic correlation effects in the new 2D materials is of paramount importance for the understanding of their transport, optical and catalytic properties. Here, by means of electron spectroscopy methods in combination with density functional theory calculations we investigate the electronic structure of a new layered van der Waals FePX$_3$ (X: S, Se) materials. Using systematic resonant photoelectron spectroscopy studies we observed strong resonant behavior for the peaks associated with the $3d^{n-1}$ final state at low binding energies for these materials. Such observations clearly assign FePX$_3$ to the class of Mott-Hubbard type insulators for which the top of the valence band is formed by the hybrid Fe-S/Se electronic states. These observations are important for the deep understanding of this new class of materials and draw perspectives for their further applications in different application areas, like (opto)spintronics and catalysis.

\end{abstract}

\section{Introduction}

The large part of research in the rapidly developing areas of science, like nanotechnology and catalysis, is devoted to the search of new materials and to the studies of their electronic structure~\cite{Ludwig.2019,Wellmann.2021}. Such experimental and theoretical studies help to make a link between fundamental knowledge and the possible applications. Here, e.\,g., layered materials, which are in the focus of the present-day intensive research, give a big hope on further development of new systems with unique electronic and transport properties, which can bring new functionalities in (nano)electronics, materials synthesis, (nano)catalysis, etc.~\cite{Kumar.20168fq,Zhan.2019,Gutierrez.2020}. The initial and following experiments on graphene, the first pure 2D material which initiated the present boom in this area, demonstrated the large perspectives of this material in the application in different areas ranging from protective layers and coatings, gas sensors to photosensors and touchscreens~\cite{Kulyk.2021,Wang.2016dhm,Larki.2020,Tkachev.2021}. However, the further progress in the implementation of graphene in the modern serial production is connected with many factors~\cite{Peplow.2015,Park.2016o1} and among them is the absence of the energy gap in the carriers' spectrum. This fact requires further modifications of graphene (chemical or structural), which in most cases lead to the uncontrollable variations of the electronic properties of the 2D layer or system~\cite{Xu.20164ud,Liu.2019h1,Yu.2020s2a,Nurazzi.2021}. The search of the new 2D materials with a natural band gap in the spectrum and the progress in their studies bring several di- and tri-atomic materials which can be considered as perspective in future applications~\cite{Manzeli.2017,Gogotsi.2019,Du.2018,Wang.20186p,Samal.2020}.

Among the layered materials which recently attract much attention are transition metal trichalcogenides (TMT) with the formula MPX$_3$ (M is the transition metal cation and X is S or Se) with large van der Waals gap between single layers~\cite{Du.2018,Samal.2020,Du.2016dn}. Because the ionic-like interaction between M$^{2+}$ and (P$_2$X$_6$)$^{4-}$ units in the single layer, these compounds in almost all cases are antiferromagnetic (AFM) wide-gap semiconductors with the energy gap ranging between $1.3$\,eV and $3.5$\,eV~\cite{Wang.20186p,Du.2016dn,Zhang.2016axgr,Yang.2020}. Due to these facts and because of the variety of the M-X combinations, these materials were proposed and tested for different applications, like photocatalytic water splitting~\cite{Du.2018,Zhang.2016axgr,Li.2019afq}, spintronic applications~\cite{Li.2013bb,Li.2014}, low-dimensional magnetic devices~\cite{Zhong.2017gqj,Pei.2017,Pei.2019qb}, and many others~\cite{Samal.2020}. Here, among other MPX$_3$, the FePX$_3$ TMTs have gained increased interest because of the recent studies of (i) magnetic properties with the observations of the giant optical linear dichroism~\cite{Zhang.2021nht,Zhang.20210x9} and (ii) photoelectrochemical water splitting where best fit of the band gap of FePX$_3$ to the redox potentials of water splitting at $pH = 0$ and $pH = 7$ was found~\cite{Zhang.2016axgr,Mukherjee.2017}. Here, Fe$^{2+}$ cations can be considered as isoelectron to Ru-ion in Ru-based complexes, which are found to be very efficient water oxidation agents~\cite{Sandhya.2011,Xue.2015,Kamdar.2019}.

Despite the widely discussed perspectives on the application of TMTs in different areas, surprisingly, not many works are devoted to the studies of the electronic structure of TMTs compounds, in particular of FePX$_3$. In most cases, only theoretical analysis of the electronic and magnetic structures is presented. However, such basic studies of the electronic structure using different spectroscopic methods, can give a direct insight in the understanding of properties which can be directly connected with the applied-oriented research. Only recently we performed intensive electronic structure studies of the layered NiPS$_3$ compound using near-edge x-ray absorption fine structure spectroscopy (NEXAFS) and resonant photoelectron spectroscopy (ResPES) which clearly identify this materials as a charge-transfer insulator~\cite{Yan.2021}. In the present work, we perform systematic combined experimental and theoretical studies of a layered FePX$_3$ (X: S, Se) family of TMTs. Our \textit{ex situ} and \textit{in situ} structural studies demonstrate very high bulk and surface quality of the studied materials permitting accurate experiments using surface-sensitive electron spectroscopy techniques, NEXAFS and ResPES. These methods, being element and orbital selective, allow in combination with density-functional theory (DFT) calculations to obtain the detailed information about the character of electronic states in the valence and conduction bands of FePX$_3$. The obtained results allow us to assign the layered FePX$_3$ materials to the Mott-Hubbard type insulator according to the Zaanen-Sawatzky-Allen scheme, opposite to the charge-transfer insulator (NiPS$_3$).

\section{Experimental details}

\paragraph*{DFT and cluster-based calculations.} 
Spin-polarized DFT calculations based on plane-wave basis sets of $500$\,eV cutoff energy were performed with the Vienna \textit{ab initio} simulation package (VASP).~\cite{Kresse.1994,Kresse.1996} The Perdew-Burke-Ernzerhof (PBE) exchange-correlation functional~\cite{Perdew.1996} was employed. The electron-ion interaction was described within the projector augmented wave (PAW) method~\cite{Blochl.1994gal} with Fe ($3p$, $3d$, $4s$), P ($3s$, $3p$), S ($3s$, $3p$) and Se ($4s$, $4p$) states treated as valence states. The Brillouin-zone integration was performed on $\Gamma$-centred symmetry reduced Monkhorst-Pack meshes using a Gaussian smearing with $\sigma = 0.05$\,eV, except for the calculation of density of states. For these calculations, the tetrahedron method with Bl\"ochl corrections~\cite{Blochl.1994} was employed. A $12\times 12\times 4$ $k$-mesh was used. The DFT+$\,U$ scheme~\cite{Anisimov.1997,Dudarev.1998} was adopted for the treatment of Fe $3d$ orbitals, with the parameter $U_\mathrm{eff}=U-J$ equal to $3$\,eV. Dispersion interactions were considered adding a $1/r^6$ atom-atom term as parameterized by Grimme (``D2'' parameterization)~\cite{Grimme.2006}. The spin-orbit interaction is taken into account. The lattice parameters of 3D FePX$_3$ ($a=b$, $c$, positions of P and S) were fully relaxed for different magnetic states. During structure optimization, the convergence criteria for energy and force were set equal to $10^{-6}$\,eV and $10^{-2}$\,eV\,\AA$^{-1}$, respectively. The obtained results listed in Table\,S1 show that both systems under study prefer the AFM state. The respective structures can be visualized using software VESTA (https://jp-minerals.org/vesta/en/).

The Fe\,$L_{2,3}$ NEXAFS spectra were calculated using CTM4XAS v5.5 program \newline(https:// anorg.chem.uu.nl/CTM4XAS/index.html)~\cite{Stavitski.2010}. The $O_h$ symmetry was considered for simplicity according to the crystallographic structure of the FePX$_3$ single layer where each X=S/Se atom is coordinated with Fe sites.

\paragraph*{Samples synthesis and characterization.} 
CVT method was used for the synthesis of FePX$_3$ crystals. Iron (99.9\%), phosphorus (99.999\%), sulphur (99.999\%) and selenium (99.999\%) from Shanghai Macklin Biochemical Co., Ltd. and Alfa Aesar were used for synthesis. A stoichiometric amount of high-purity elements (mole ratio Ni : P : S/Se = 1 : 1 : 3, 1\,g in total) and iodine (about $20$\,mg) as a transport agent were sealed into a quartz ampule (length is $15$\,cm, external diameter is approximately $15$\,mm) and kept in a two-zone furnace with $700-650^\circ$\,C and $650-600^\circ$\,C for FePS$_3$ and FePSe$_3$, respectively. The vacuum inside the ampule was pumped down to $1\times10^{-3}$\,mbar. After 7 days of heating, the ampule was cooled down to room temperature with bulk crystals in the colder edge.

After synthesis, the systematic characterization of samples was performed: (i) optical images were obtained with optical microscope at different magnifications; (ii) XRD patterns were collected at room temperature with a Bruker D2 Phaser diffractometer using Cu\, $K$ ($1.54178$\,\AA) radiation; (iii) for the Raman characterization, a Nanofinder 30 was used to obtain high-resolution spectra. The FePX$_3$ samples were illuminated with a laser wavelength of $532$\,nm and power $32$\,mW with a $1\,\mu\mathrm{m}$ laser spot.  Before characterization, standard single-crystal silicon was used to calibrate the system. For measurements, an integration time (the exposure time) of $5$\,s and an accumulation of $3$ were set to obtain spectra with high signal-to-noise ratio; (iv) TEM measurements were performed using FEI Tecnai G2 F30 instrument and the focused ion beam preparation was performed using FEI Strata 400S; (v) SEM/EDX data were collected using ZEISS SIGMA 500 microscope; (vi) LEED measurements were performed in ultra-high vacuum (UHV) conditions using MCP-LEED (Scienta Omicron) with the electron beam current of $50-200$\,pA.

\paragraph*{PES and NEXAFS experiments.}
All photoelectron spectroscopy measurements were performed at the Russian-German dipole soft x-ray beamline (RGBL-dipole) of the BESSY\,II synchrotron radiation facility (Helmholtz-Zentrum Berlin)~\cite{Molodtsov.2009}. All spectra were measured in UHV conditions (vacuum is below than $2\times10^{-10}$\,mbar) and at room temperature. Before every set of experiments, new sample was cleaved in air and immediately introduced in vacuum, followed by the degassing routine at $350^\circ$\,C for $60$\,min. NEXAFS spectra were collected in the total electron yield (TEY) mode using sample drain current. XPS core-level and ResPES valence band spectra were acquired using SPECS PHOIBOS 150 analyser. Photon energies in all experiments were calibrated using a polycrystalline Au sample. During ResPES measurements the effect of a sample charging was minimal (within $1$\,eV) and was not compensated; however, the correctness of every measurement was confirmed by the acquisition of the respective S\,$2p$ or Se\,$3p$ spectra at every photon energy step before and after collection of the corresponding valence band spectra.

\section{Results and Discussion}

\begin{figure}
\includegraphics[width=\textwidth]{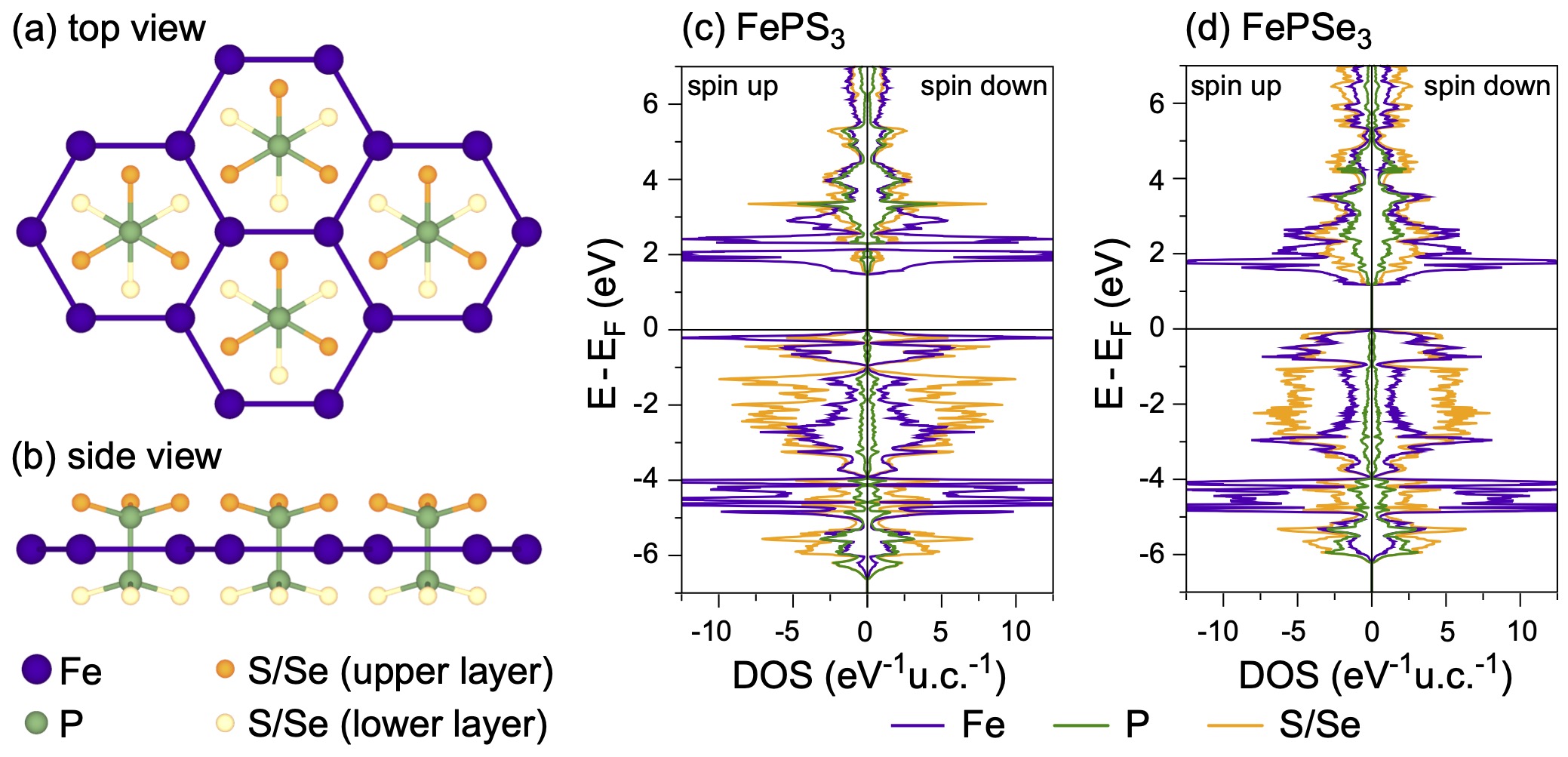}
\caption{\label{fig:FePX3_structure_DOS} Top (a) and side (b) views of the crystallographic structure of the FePX$_3$ (X: S or Se) single layer. Atom-projected partial DOS for bulk FePS$_3$ (c) and FePSe$_3$ (d) in the AFM ground state.}
\end{figure}

\textit{DFT.} The crystallographic structure of a FePX$_3$ monolayer can be considered as a layer of MoS$_2$ where one third of metal atoms is replaced by the P--P dimers which are perpendicular to the TMT layer and form the ethane-like structure with X-atoms (Fig.~\ref{fig:FePX3_structure_DOS}(a,b)), thus Fe-cations form the graphene-like honeycomb structure. The monolayers of FePX$_3$ stack in the $C2/m$ and $R\overline{3}$ space groups with the van der Waals gap of $3.32$\,\AA\ and $3.26$\,\AA\ between neighboring chalcogen layers for X\,=\,S and X\,=\,Se, respectively (see Fig.\,S1, Tab.\,S1 and corresponding structural files in the Supplementary Information). Both 3D FePX$_3$ compounds are AFM in the ground states with the distance between layers $c'=6.537 (6.621)$\,\AA\ for X\,=\,S\,(Se), respectively. The mixing character of interaction (ionic and covalent) in the structural unit of FePX$_3$ leads to the specific distribution of valence band states in the calculated density of states (DOS). As can be seen, both FePX$_3$ compounds are wide band gap semiconductors with an energy gap of $1.48$\,eV and $1.17$\,eV for X\,=\,S and X\,=\,Se, respectively (Fig.~\ref{fig:FePX3_structure_DOS}(c,d)). The space and energy overlap between Fe\,$3d$ and S/Se,$sp$ leads to the formation of the respective hybrid bands in the whole energy range of DOS. The top of the valence band of FePX$_3$ is formed by the hybrid Fe-S/Se states with the significant contribution of the Fe\,$3d$ partial DOS. The same is also valid for the bottom of the conduction band. At the same time the less localized character of the electronic states, and particularly of the Fe\,$3d$ states, can be noted for FePSe$_3$.

The presented distribution of the electronic states in DOS of FePX$_3$ is different from those for isostructural NiPS$_3$ and similar to the states distribution for isostructural MnPX$_3$ (although not so energy localized for the metal-projected partial DOS)~\cite{Yang.2020,Yan.2021,Sen.2020da}. According to the Zaanen-Sawatzky-Allen scheme~\cite{Zaanen.1985}, the electronic structure of oxides and sulphides can be described with two parameters, the $d-d$ correlation energy $(U_{dd})$ and charge transfer energy between $d$-states of the metal and $p$-states of the ligand $(\Delta)$. Taking into account this scheme, the later compounds can be assigned to two limit cases of Mott-Hubbard type insulator $(U_{dd} < \Delta)$ for MnPX$_3$ (Mn$^{2+}\,3d^5$) and charge-transfer type insulator $(U_{dd} > \Delta)$ for NiPS$_3$ (Ni$^{2+}\,3d^8$). In case of FePX$_3$ the obtained values of $U_{dd}=3$\,eV and $\Delta\approx6$\,eV ($\Delta$ is taken as a difference between centres of gravity for the occupied and unoccupied ligand partial DOS) place these materials to the case of Mott-Hubbard type insulator.

\begin{figure}
\includegraphics[width=\textwidth]{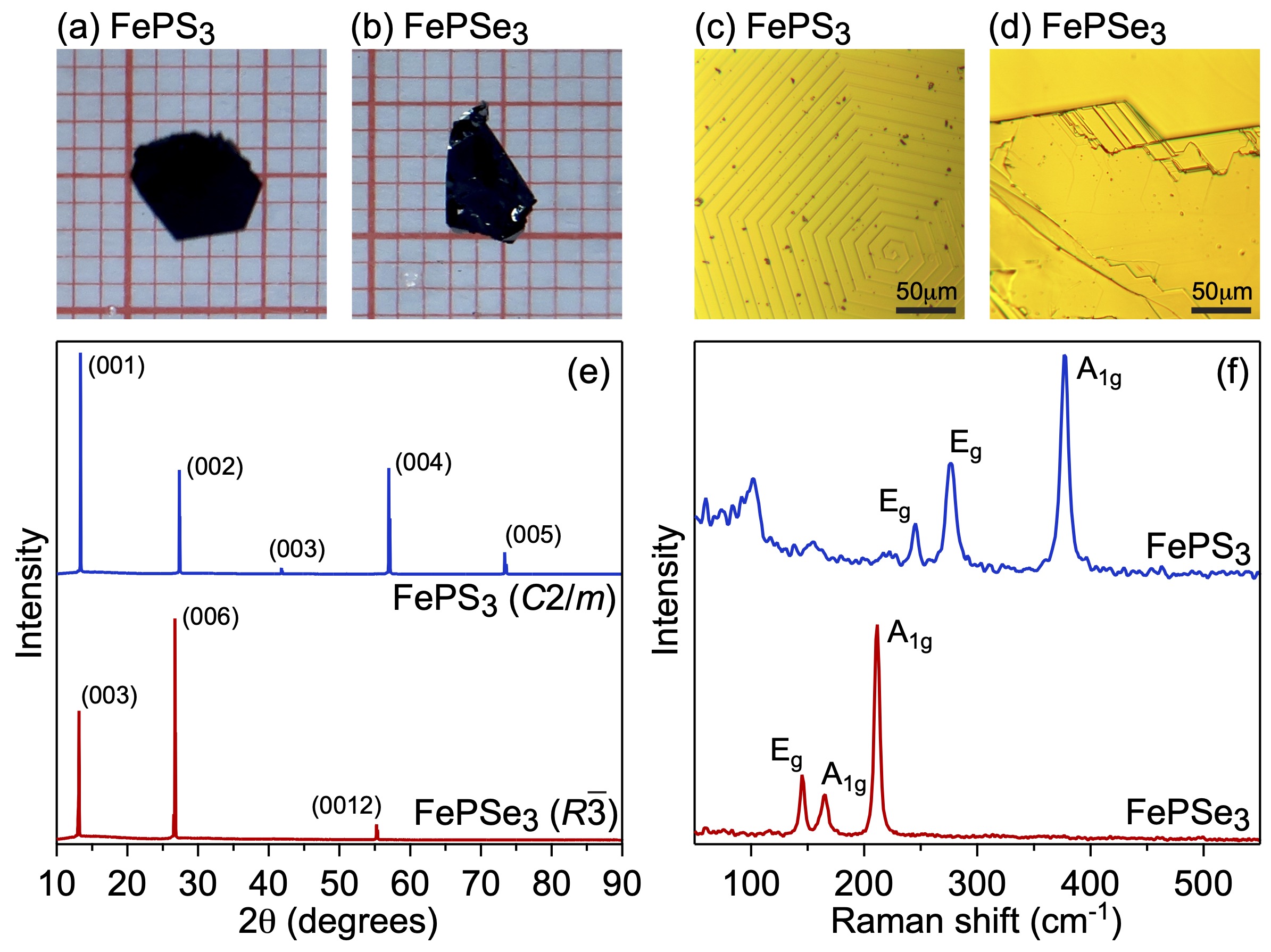}
\caption{\label{fig:FePX3_Photo_XRD_Raman} Characterization of FePX$_3$ crystals: (a,b) photos, (c,d) optical microscopy images, (e) XRD patterns and (f) Raman spectra measured at room temperature for bulk samples.}
\end{figure}

\textit{Characterization.} Figures~\ref{fig:FePX3_Photo_XRD_Raman} and S2 summarize the experimental results on the bulk characterization of the FePX$_3$ crystals studied in the present work. Well-ordered single crystalline samples with the linear sizes up to several-mm clearly demonstrate layered structure in optical images with angles between crystalline edges either $60^\circ$ or $120^\circ$ confirming hexagonal atomic arrangement in a single FePX$_3$ layer (Fig.~\ref{fig:FePX3_Photo_XRD_Raman}(a-d)). In the XRD plots of the FePX$_3$ layered samples there are only $(00l)$ preferred orientation peaks which confirm high bulk quality and the respective space symmetry groups of FePX$_3$ crystals (Fig.~\ref{fig:FePX3_Photo_XRD_Raman}(e)) \cite{Du.2016dn,Xie.20191k}. The extracted from the TEM data the interlayer distances for FePS$_3$ and FePSe$_3$ are $6.45$\,\AA\ and $6.64$\,\AA, respectively (Fig.\,S2), which are in very good agreement with previously published structural data for these compounds~\cite{Du.2016dn} and with our present theoretical results (see above). Raman spectroscopy characterization also confirm a high quality of the studied FePX$_3$ crystals where observed peaks can be assigned to the corresponding vibrations involving metal atoms (broad bands at $101\,\mathrm{cm}^{-1}$ for FePS$_3$ and $117\,\mathrm{cm}^{-1}$ for FePSe$_3$) and the other come from the vibrations of the [P$_2$X$_6$] unit with a $D_{3d}$ symmetry group (FePS$_3$: $155\,\mathrm{cm}^{-1}$, $245\,\mathrm{cm}^{-1}$, $276\,\mathrm{cm}^{-1}$, $376\,\mathrm{cm}^{-1}$; FePSe$_3$: $145\,\mathrm{cm}^{-1}$, $165\,\mathrm{cm}^{-1}$, $211.5\,\mathrm{cm}^{-1}$)~\cite{Wang.2016udy,Du.2016dn,Mukherjee.2016,Mukherjee.2017}.

\begin{figure}
\includegraphics[width=\textwidth]{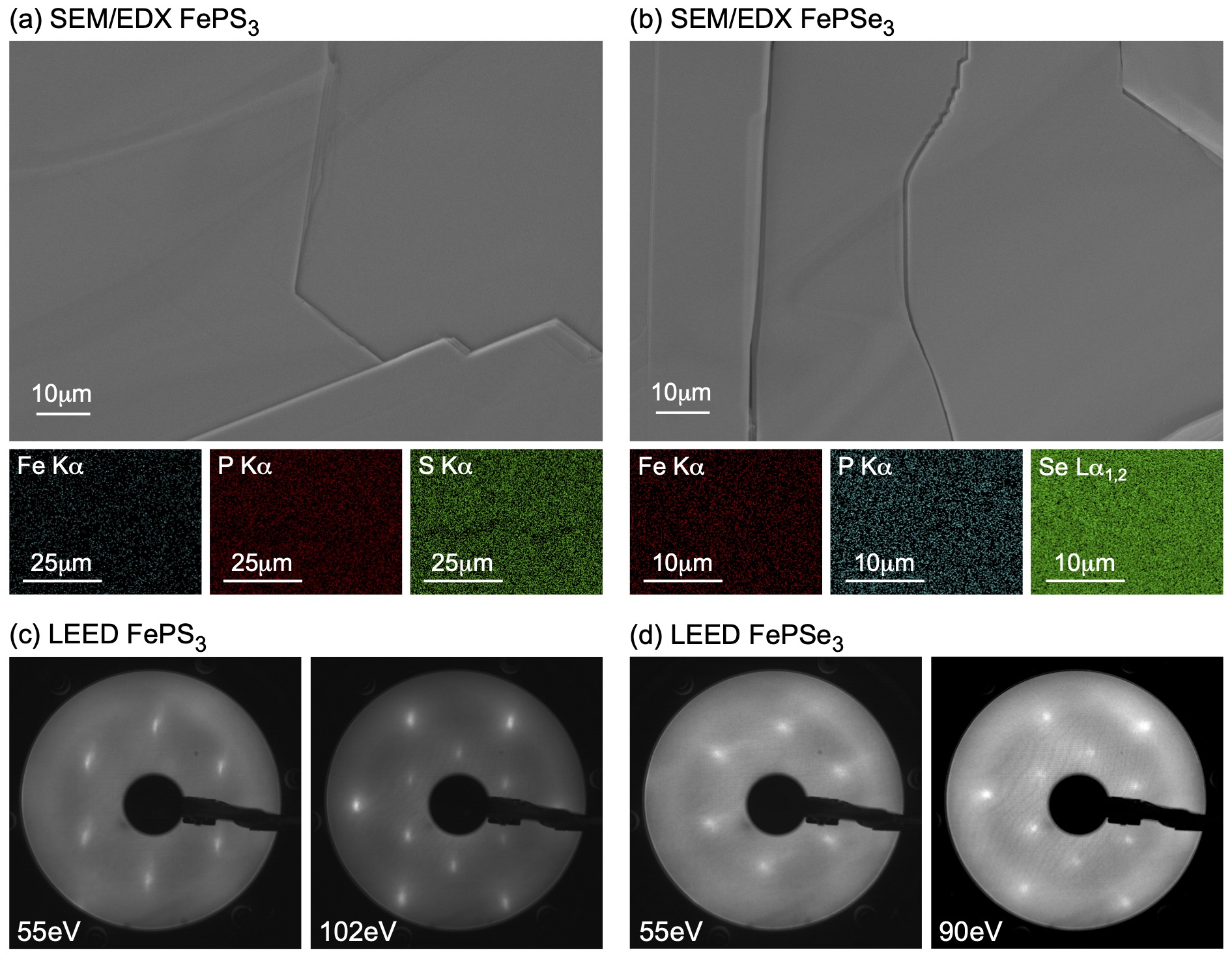} 
\caption{\label{fig:FePX3_SEM_LEED} (a,b) SEM and the respective EDX maps collected at the primary electron beam energy of $15$\,keV for FePS$_3$ and FePSe$_3$, respectively. (c,d) Set of LEED images collected for freshly cleaved and annealed FePS$_3$ and FePSe$_3$ samples, respectively. Energies for the primary electron beam are marked in every image.}
\end{figure}

Freshly cleaved surfaces of FePX$_3$ crystals were used for SEM/EDX measurements (Fig.~\ref{fig:FePX3_SEM_LEED}(a,b)). The obtained SEM images also confirm the high crystallographic quality of the obtained surfaces with the terraces width of several hundreds $\mu\mathrm{m}$. The measured EDX maps shown below the respective SEM images confirm the uniform distribution of elements and the stoichiometry of the studied samples (see Fig.\,S3). After characterization, freshly cleaved samples were introduced in UHV and annealed at $350^\circ$\,C. Following this procedure, the measured LEED images (Fig.~\ref{fig:FePX3_SEM_LEED}(c,d)) show clear hexagonal diffraction patterns indicating the long-range structural ordering of the FePX$_3$(001) surfaces without structural defects and surface adsorbates. The slightly diffuse diffraction spots observed at low primary electron beam energies can be assigned to the residual charging effects for the wide-band semiconducting FePX$_3$ and the less sharp picture in case of FePSe$_3$(001) is due to the less structural ordering of the FePSe$_3$ samples as can be seen from optical and SEM images.

\begin{figure}
\includegraphics[width=\textwidth]{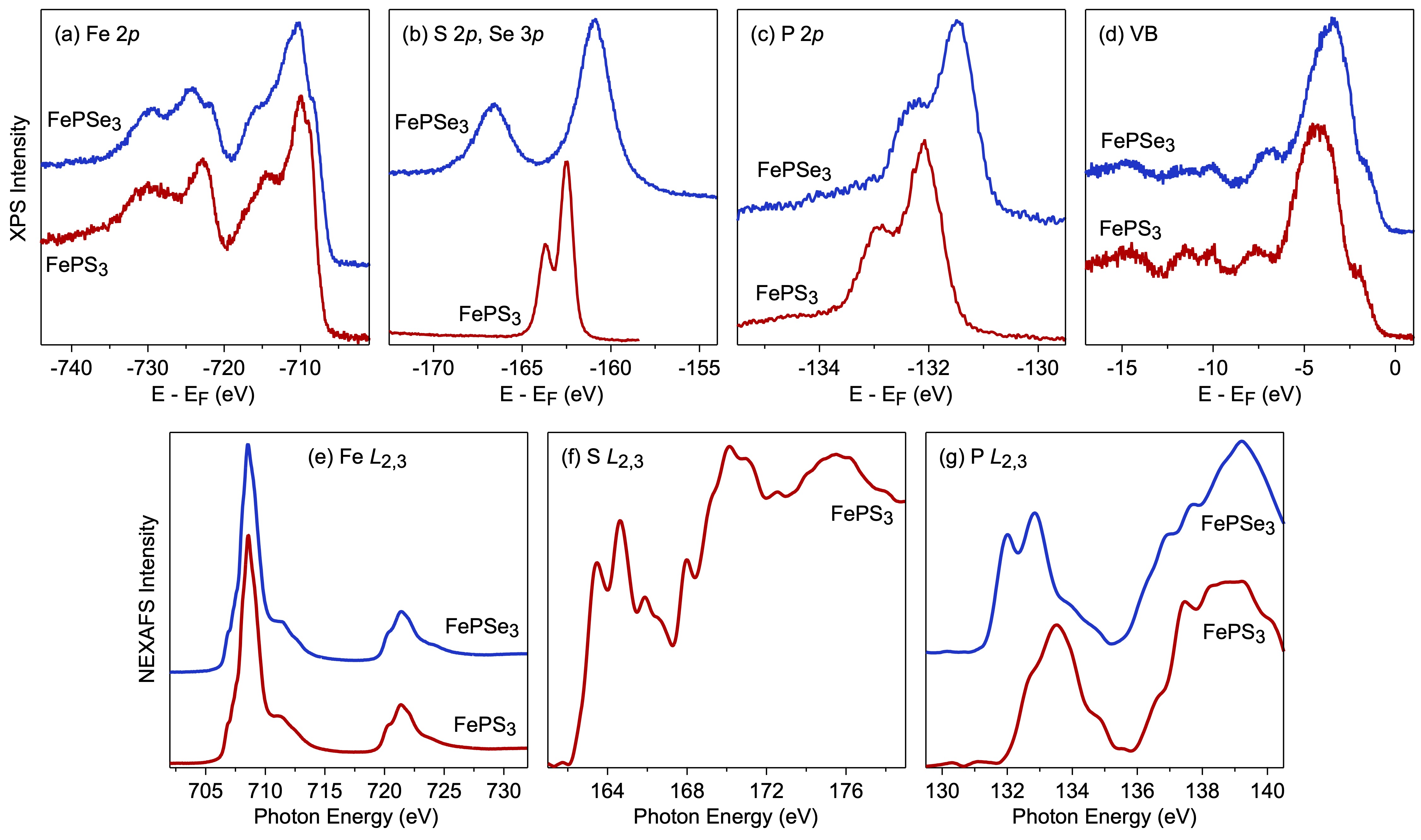}
\caption{\label{fig:FePX3_XPS_NEXAFS} XPS spectra of FePX$_3$ collected at $h\nu = 1000$\,eV: (a) Fe\,$2p$, (b) S\,$2p$ and Se\,$3p$, (c) P\,$2p$, and (d) valence band region. NEXAFS spectra of FePX$_3$ collected in the TEY mode: (e) Fe\,$L_{2,3}$, (f) S\,$L_{2,3}$, and (g) P\,$L_{2,3}$.
}
\end{figure}

\textit{XPS and NEXAFS.} Figure~\ref{fig:FePX3_XPS_NEXAFS} shows compilation of the XPS and NEXAFS spectra of freshly-cleaved and UHV-annealed FePX$_3$ crystals collected for the representative core-level emission lines and at the respective absorption edges (the Se\,$M_{4,5}$ spectrum for FePSe$_3$ is not shown). In these compounds Fe$^{2+}$ cations are octahedrally coordinated with S-ions. Therefore Fe\,$2p$ XPS (Fig.~\ref{fig:FePX3_XPS_NEXAFS}(a)) and Fe\,$L_{2,3}$ NEXAFS spectra (Fig.~\ref{fig:FePX3_XPS_NEXAFS}(e)) are very similar to those for FeO (Fe$^{2+}$), which is assigned to the class of Mott-Hubbard insulator~\cite{Gota.1999,Yamashita.2008,Miedema.20138et,Ahmed.2012,Liu.2019nhq}. Particularly it is visible for the photoemission satellite structure for the respective XPS emission line. The extracted energy difference between Fe\,$2p_{3/2}$ main line and the satellite is $4.5$\,eV and $4.95$\,eV for FePS$_3$ and FePSe$_3$, respectively. These values can be compared to the corresponding ones of $5.5$\,eV for FeO (Fe$^{2+}$) and $8.0$\,eV for Fe$_2$O$_3$ (Fe$^{3+}$)~\cite{Gota.1999,Yamashita.2008}. The interesting feature of the Fe\,$2p$ XPS spectra for FePX$_3$ materials is the observation of the ``\textit{metallic}-like'' feature located at $708.5$\,eV ($721.9$\,eV) of binding energy for Fe\,$2p_{3/2}$ (Fe\,$2p_{1/2}$) spin-orbit split components (it is mostly pronounced for FePSe$_3$) and which was not observed for the previously studied isostructural NiPS$_3$~\cite{Yan.2021}. The nature of this spectral feature is not fully clear and could be a subject of further studies.

The same consideration is also valid for the description of the Fe\,$L_{2,3}$ NEXAFS spectra of FePX$_3$ (the respective NEXAFS spectra for both compounds are almost identical) (Fig.~\ref{fig:FePX3_XPS_NEXAFS}(e)). Generally, in case of FeO and FePX$_3$ for the Fe$^{2+}$ cation in the octahedral ligand field the Fe\,$L_{2,3}$ NEXAFS spectra are very similar to each other (cf. Fig.~\ref{fig:FePX3_XPS_NEXAFS}(e) and data in Refs.~\cite{Miedema.20138et,Ahmed.2012,Yang.2012j2e}) and different from the one for Fe$_3$O$_4$ (Fe$^{2+}$/Fe$^{3+}$) and Fe$_2$O$_3$ (Fe$^{3+}$). According to the crystal-field approach, the NEXAFS spectra of FeO were simulated using the octahedral crystal field splitting parameter $10Dq=0.4$\,eV between $t_{2g}$ and $e_g$ orbitals~\cite{Miedema.20138et}. In the present modelling using crystal-filed approach we used slightly smaller value of $10Dq=0.3$\,eV due to the weaker ligand field of the S/Se ions and the good agreement with the experimental spectra is found for the values of $U_{dd}=3$\,eV and $\Delta=6$\,eV (Fig.\,S4) confirming the description of FePX$_3$ as a Mott-Hubbard type insulator.

The S\,$2p$/Se\,$3p$ and P\,$2p$ XPS spectra of FePX$_3$ show clear spin-orbit split structure (Fig.~\ref{fig:FePX3_XPS_NEXAFS}(b,c)). The difference in the binding energy of the P\,$2p$ line between FePS$_3$ and FePSe$_3$ of $\approx0.6$\,eV is due to the slightly smaller electronegativity of Se compared to S and to the smaller value of the valence band maximum (VBM) position extracted from the valence band spectra -- $E_\mathrm{VBM}-E_F=-0.78$\,eV for FePS$_3$ vs. $E_\mathrm{VBM}-E_F=-0.45$\,eV for FePSe$_3$ (because of the different band gaps of two materials; see above) (Fig.~\ref{fig:FePX3_XPS_NEXAFS}(d)). The respective S\,$L_{2,3}$ and P\,$L_{2,3}$ NEXAFS spectra of FePX$_3$ (Fig.~\ref{fig:FePX3_XPS_NEXAFS}(f,g)) correspond to the electronic transitions from the S/P spin-orbit split $2p_{3/2,1/2}$ level onto the unoccupied $d$ and $s$ states in the conduction band. As was stated in Ref.~\cite{Yan.2021}, the interpretation of these spectra is not a trivial task and, as was already discussed, the structure in the S\,$L_{2,3}$ and in the P\,$L_{2,3}$ spectra at $\approx161-167$\,eV and $\approx131-136$\,eV of photon energy, respectively, can be assigned to the electron transitions into the first unoccupied hybrid $3s$-like antibonding state formed by S/Se and P~\cite{Farrell.2002eb,Kruse.2009,Yang.2012j2e,Yan.2021}. Additional high photon energy structure in the NEXAFS spectra can be assigned to the so-called ``echo'' or shadow effect of the spin-orbit split features due to multiple scattering or to electron transitions to a mixed-valence band states~\cite{Farrell.2002eb}.

\begin{figure}
\includegraphics[width=0.75\textwidth]{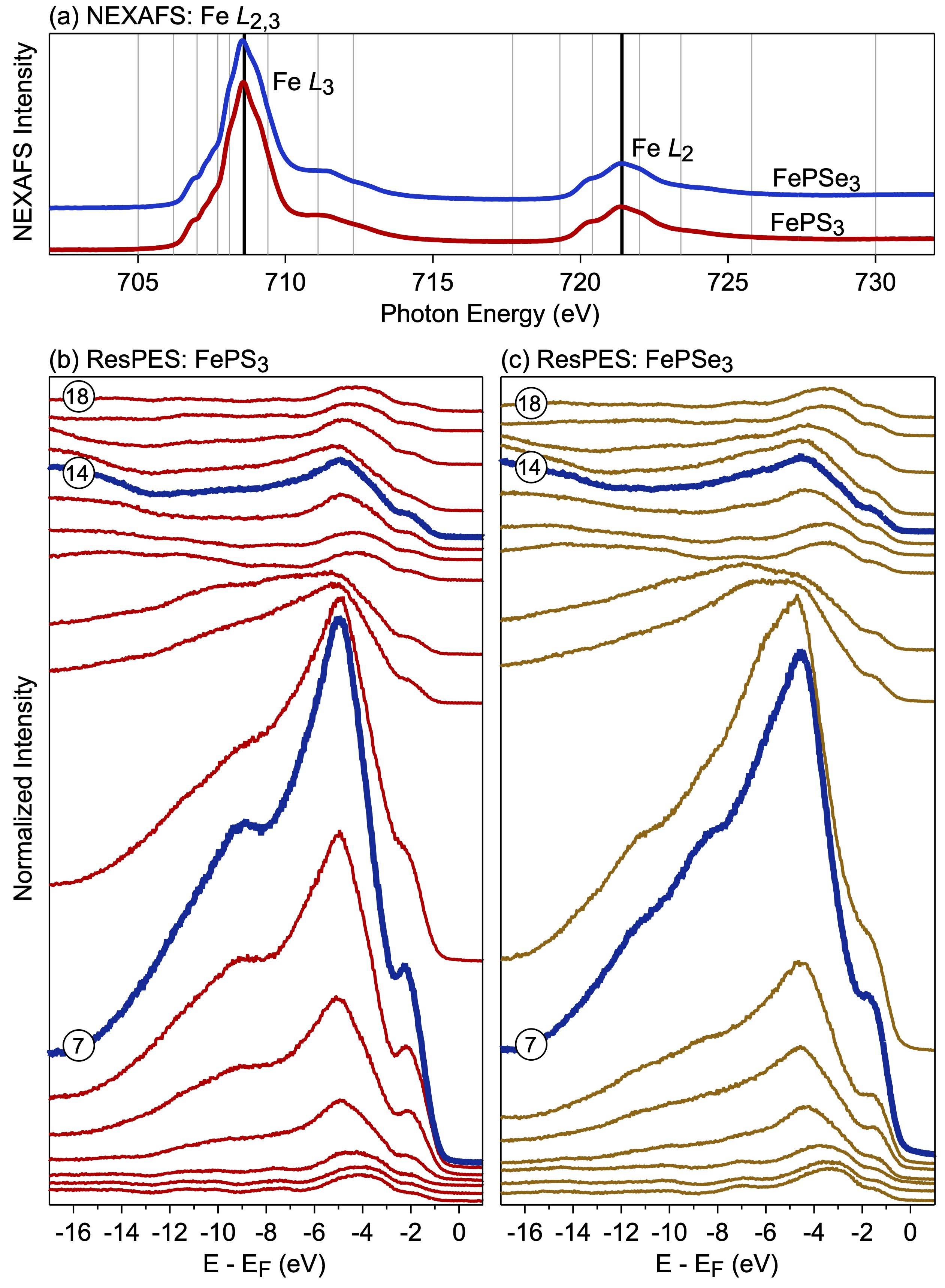}
\caption{\label{fig:FePX3_ResPES} (a) The reference Fe\,$L_{2,3}$ NEXAFS spectra and (b,c) a series of photoemission spectra taken at the particular photon energies marked by the corresponding vertical line in panel (a) for FePX$_3$ crystals. All ResPES spectra are shifted in the vertical direction for clarity.}
\end{figure}

\textit{ResPES at the Fe\,$L_{2,3}$ edge.} The method of the resonant photoelectron spectroscopy for electronic correlated systems is a unique tool allowing to clearly assign the studied objects to one of the classes of insulating systems which are classified according to the discussed Zaanen-Sawatzky-Allen scheme~\cite{Zaanen.1985}. Figure~\ref{fig:FePX3_ResPES} presents the compilation of the respective results for FePX$_3$ crystals: (a) reference Fe\,$L_{2,3}$ NEXAFS spectra and (b,c) a series of photoelectron emission spectra collected at the particular photon energies marked by the vertical lines in panel (a). For the ResPES measurements of $3d$-derived valence band states at the Fe\,$L_{2,3}$ absorption edge on FePX$_3$, the photoemission intensity is a result of the interference of two photoemission channels~\cite{Davis.1981abc,Davis.1982,Solomon.2005}: (i) a direct photoelectron emission from the valence band states $2p^63d^n + h\nu \rightarrow 2p^63d^{n-1} + e$ and (ii) a photoabsorption process followed by a participator Coster-Kronig decay $2p^63d^n + h\nu \rightarrow 2p^53d^{n+1} \rightarrow 2p^63d^{n-1} + e$, where final states for these photoelectron emission channels are identical. The interference between these two photoemission channels leads to the Fano-type resonance for the states with the $3d^{n-1}$ final-state character.

For transition metal chalcogenides (oxides or sulfides) it was shown~\cite{Fujimori.1990,Okada.1992,Tjernberg.1996,Nesbitt.2000,Taguchi.2007} that in case of the charge-transfer insulator state with $U_{dd} > \Delta$ (NiO, NiS, or NiS$_2$) the core-level or valence band XPS spectra consist of the intense low binding energy peak which is formed by the mixing of the $3d^n\underline{L}$ and $3d^n\underline{Z}$ final states ($\underline{L}$ and $\underline{Z}$ are a ligand hole and Zhang-Rice doublet bound states, respectively), and the high binding energy peak of the lower intensity is connected with the $3d^{n-1}$ final state. In case of the Mott-Hubbard type insulator state when $U_{dd} < \Delta$ (MnO or FeO) the situation is reversed and the intense low binding energy peak originates mainly from the $3d^{n-1}$ final state with the low intensity satellite $3d^n\underline{L}$ structure at large binding energies. Therefore, taking into account the fact that for the $2p \rightarrow 3d$ resonant photoemission the Coster-Kronig decay channel (marked as (ii)) dominates in the photoemission spectra~\cite{Davis.1981abc,Davis.1982}, these measurements can give a position of the $3d^{n-1}$ peak relative to the one for the $3d^n\underline{L}$ allowing to assign the measured material to the particular insulator class. Recently, such experiments and analysis of the ResPES data allowed to identify the isostructural layered NiPS$_3$ to the charge transfer insulator state~\cite{Yan.2021}. (Here we have to note that the clear separation between two final states is not possible due to the strong hybridization between them requiring the detailed analysis of the measured resonance photoemission spectra using cluster model calculations.)

In Fig.~\ref{fig:FePX3_ResPES}(b,c) the off-resonance spectra of FePX$_3$ collected at $h\nu=705$\,eV demonstrate a broad band in the range of $E-E_F\approx-2.5\dots-6.5$\,eV with a shoulder towards $E_F$ (see also Fig.~\ref{fig:FePX3_XPS_NEXAFS}(d)). A set of photoemission satellites is observed at binding energies $E-E_F<-7$\,eV. According to our previous theoretical and NEXAFS data, FePX$_3$ are identified as Mott-Hubbard type insulators with $U_{dd} < \Delta$; therefore, the low binding energy structure in the valence band XPS spectra can be assigned to the $3d^{n-1}$ final state structure and the satellite structure to the $3d^n\underline{L}$ final state. The on-resonance spectrum (spectra 7 in Fig.~\ref{fig:FePX3_ResPES}(b,c)) collected at the photon energy of $h\nu=708.6$\,eV (Fe\,$L_3$ absorption edge) demonstrates the drastic increase of the photoemission intensity by factor of $\approx25$ in the low binding energy range of $E-E_F\approx0\dots-6.5$ (with additional shoulders at $E-E_F\approx-8.5$\,eV and $E-E_F\approx-10.7$\,eV). Taking into account the previous considerations of the ResPES process and that $3d^{n-1}$ Coster-Kronig decay channel dominates the photoemission intensity, we can confirm that the low binding energy resonating structure in the photoemission spectra correspond to the $3d^{n-1}$ final state strongly supporting the description of the FePX$_3$ materials as a Mott-Hubbard type insulator. The similar consideration is also valid for the resonating behavior of the photoemission spectra at the Fe\,$L_2$ absorption edge (spectra 14 in Fig.~\ref{fig:FePX3_ResPES}(b,c)). The presented results are in a rather good agreement with the calculated ResPES spectra of FeO (Fe$^{2+}$) which is also described as a Mott-Hubbard type insulator~\cite{Tanaka.1994}. Our results are also supported by the presented DFT calculations for FePX$_3$ (see Fig.~\ref{fig:FePX3_structure_DOS}(c,d)). As was discussed earlier, the top of the valence band is formed by the hybrid Fe($3d$)-S($3p$)/Se($4p$) states with significant contribution of the Fe-derived states. At the same time the bottom of the conduction band is formed by mainly Fe($3d$) states, thus confirming the Mott insulator state for FePX$_3$. Also, the more localized nature of the Fe\,$3d$ valence band states in the vicinity of $E_F$ for FePS$_3$ compared to FePSe$_3$ (see Fig.~\ref{fig:FePX3_structure_DOS}(c,d)) is experimentally confirmed by the observation of the sharper photoemission intensity of the low binding energy feature at $E-E_F\approx-2.25$\,eV for FePS$_3$ (Fig.~\ref{fig:FePX3_ResPES}(b,c)).

\section{Conclusion}

In summary, the crystallographic structure and electronic properties of high-quality layered FePX$_3$ (X: S, Se) crystals were studied using different experimental methods including \textit{in situ} UHV surface science techniques, like LEED, NEXAFS, XPS and ResPES in combination with systematic DFT calculations. Our theoretical and spectroscopy results indicate the strong correlation effects in the electronic structure of FePX$_3$ and using NEXAFS, XPS and ResPES methods it is shown that these materials can be described as Mott-Hubbard type insulator with $U_{dd} < \Delta$ according to the Zaanen-Sawatzky-Allen scheme. Particularly, the systematic ResPES experiments performed around the Fe\,$L_{2,3}$ absorption edge show the strong resonant behavior for the $3d^{n-1}$ final states located at low binding energies in the electronic structure of FePX$_3$. Such behavior is opposite to the case of the isostructural layered NiPS$_3$ materials which was identified as a charge transfer insulator according to the same classification scheme. The present theoretical and spectroscopic results are of a great importance for the fundamental investigations and understanding of the new class of layered materials - transition-metal trichalcogenides, as they provide the clear description of their electronic structure, which is important for further understanding of their transport, optical and catalytic properties. Particularly, it can be important for the understanding of the reaction of FePX$_3$ towards water splitting, where Fe\,$3d$ states located at the top of the valence band might play a crucial role in this process, bringing these materials as perspective for the low-dimensional oxygen- and hydrogen-gas generation.

\begin{acknowledgement}
We would like to acknowledge Anna Makarova and Dmitry Smirnov for their technical support during beamtime at HZB. Y.D. and E.V. thank the ``ENSEMBLE3 - Centre of Excellence for nanophotonics, advanced materials and novel crystal growth-based technologies'' project (GA No. MAB/2020/14) carried out within the International Research Agendas programme of the Foundation for Polish Science co-financed by the European Union under the European Regional Development Fund and the European Union's Horizon 2020 research and innovation programme Teaming for Excellence (GA. No. 857543) for support of this work. We thank HZB for the allocation of synchrotron radiation beamtime and for support within the bilateral Russian-German Laboratory program. The North-German Supercomputing Alliance (HLRN) is acknowledged for providing computer time.
\end{acknowledgement}

\begin{suppinfo}
The following files are available free of charge.
\begin{itemize}
  \item Additional experimental and theoretical data (PDF)
  \item Structural data for the 3D bulk FePS$_3$ (TXT)
   \item Structural data for the 3D bulk FePSe$_3$ (TXT)
\end{itemize}

\end{suppinfo}



\clearpage

\clearpage

\clearpage

\clearpage

\clearpage

\clearpage
\noindent
\textbf
{Supplementary Information for Manuscript: ``Mott-Hubbard Insulating State for the Layered van der Waals FePX$_3$ (X:S, Se) As Revealed by NEXAFS and Resonant Photoelectron Spectroscopy''}




\clearpage
\begin{figure}
\centering
\includegraphics[width=0.4\textwidth]{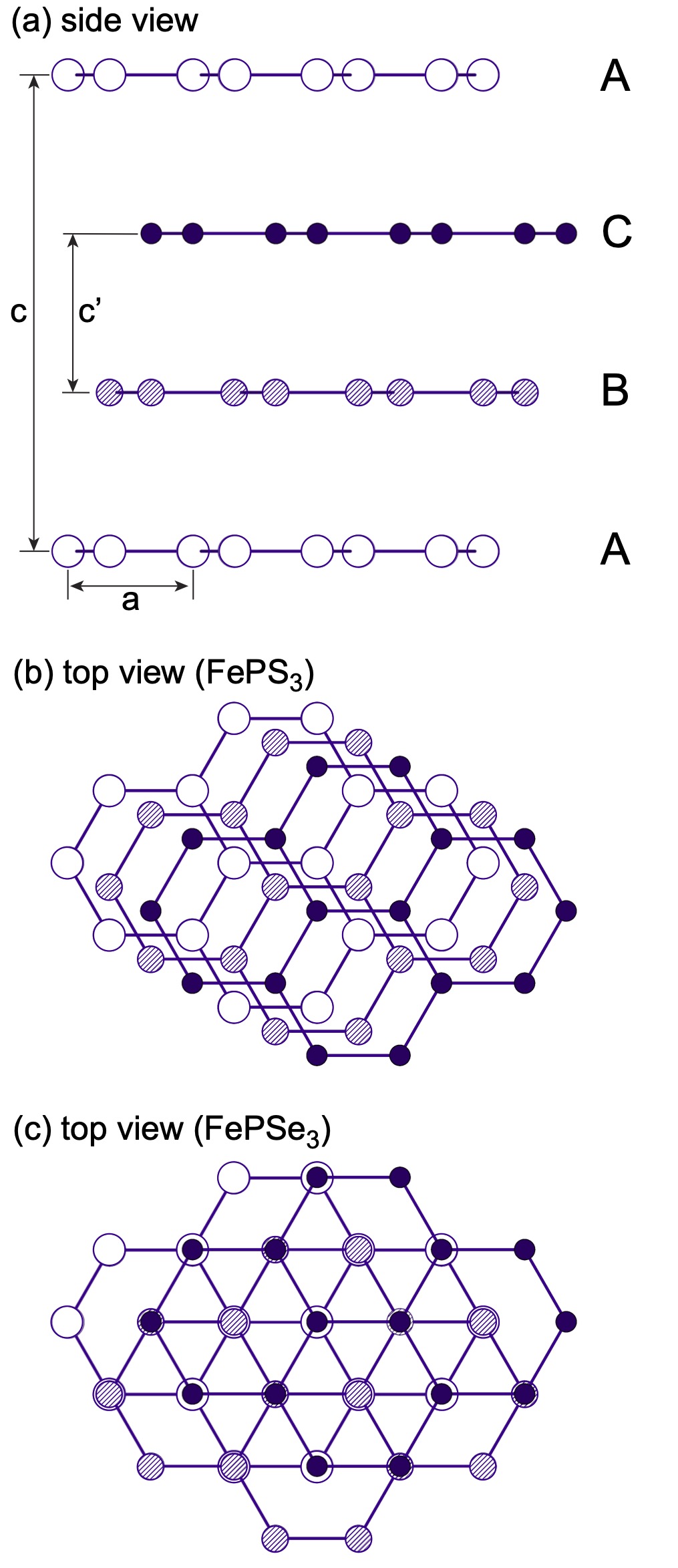}
\end{figure}
\noindent Fig.\,S1. $ABC$-sequence of layers in 3D FePX$_3$. For simplicity, only Fe-ions are shown. Atoms of different layers are shown with spheres of different size and style. In (a) the in-plane and out-of-plane lattice constants are indicated with letters a and c, respectively; $c'$$=$$c/3$ is the distance between single layers; (b) and (c) highlight the difference between stackings in FePS$_3$ and FePSe$_3$.

\clearpage
\begin{table}
\centering
\medskip
	\setlength{\tabcolsep}{1.5em}
	\begin{tabular}{cccrccc}
		\hline
System &	Magn. state&$E$& $\Delta E$ &$a$ & $c$ & $c'$\\
\hline
FePS$_3$	&AFM	&$-160.799$ &$0$		& $5.936$ &$19.611$ & $6.537$\\
			&FM   	&$-160.646$ &$153$	& $5.948$ &$19.650$ &$6.550$\\
         		&NM   	&$-154.315$ &$6485$ 	& $5.720$ &$18.808$& $6.269$\\[0.2cm]
FePSe$_3$ 	&AFM  	&$-147.449$ &$0$   		& $6.279$ &$19.863$& $6.621$\\
			&FM   	&$-147.357$ &$92$  	& $6.279$ &$19.865$& $6.622$\\
			&NM   	&$-141.099$ &$6350$ 	& $6.071$ &$19.260$& $6.420$\\
\hline
\end{tabular} 
\end{table}
\noindent Tab.\,S1. Results for the 3D bulk structure of FePX$_3$ obtained for different magnetic states: $E$ (in eV) is the total energy; $\Delta E$ (in meV) is the energy difference between the energy calculated for the different magnetic states and the energy calculated for the lowest energy structure; $a$, $c$ (in \AA) are the in-plane and out-of-plane lattice constants; $c'$$=$$c/3$ (in \AA) is the distance between single layers (cf. Figure~S1).

\clearpage
\begin{figure}
\includegraphics[width=\textwidth]{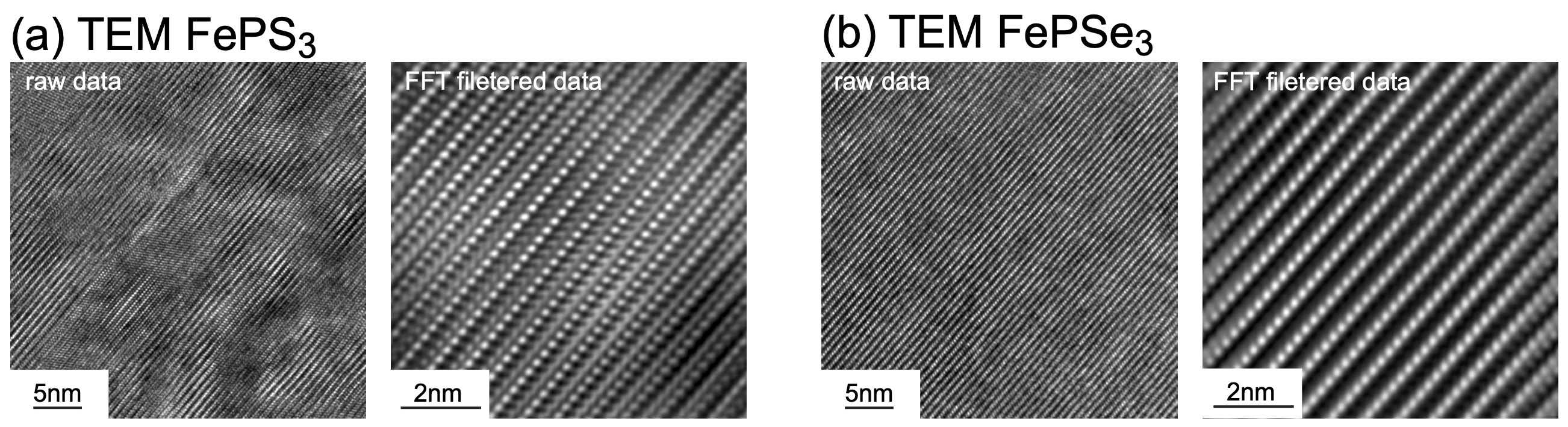}
\end{figure}
\noindent Fig.\,S2. High-resolution TEM images of bulk (a) FePS$_3$ and (b) FePSe$_3$.

\clearpage
\begin{figure}
\includegraphics[width=\textwidth]{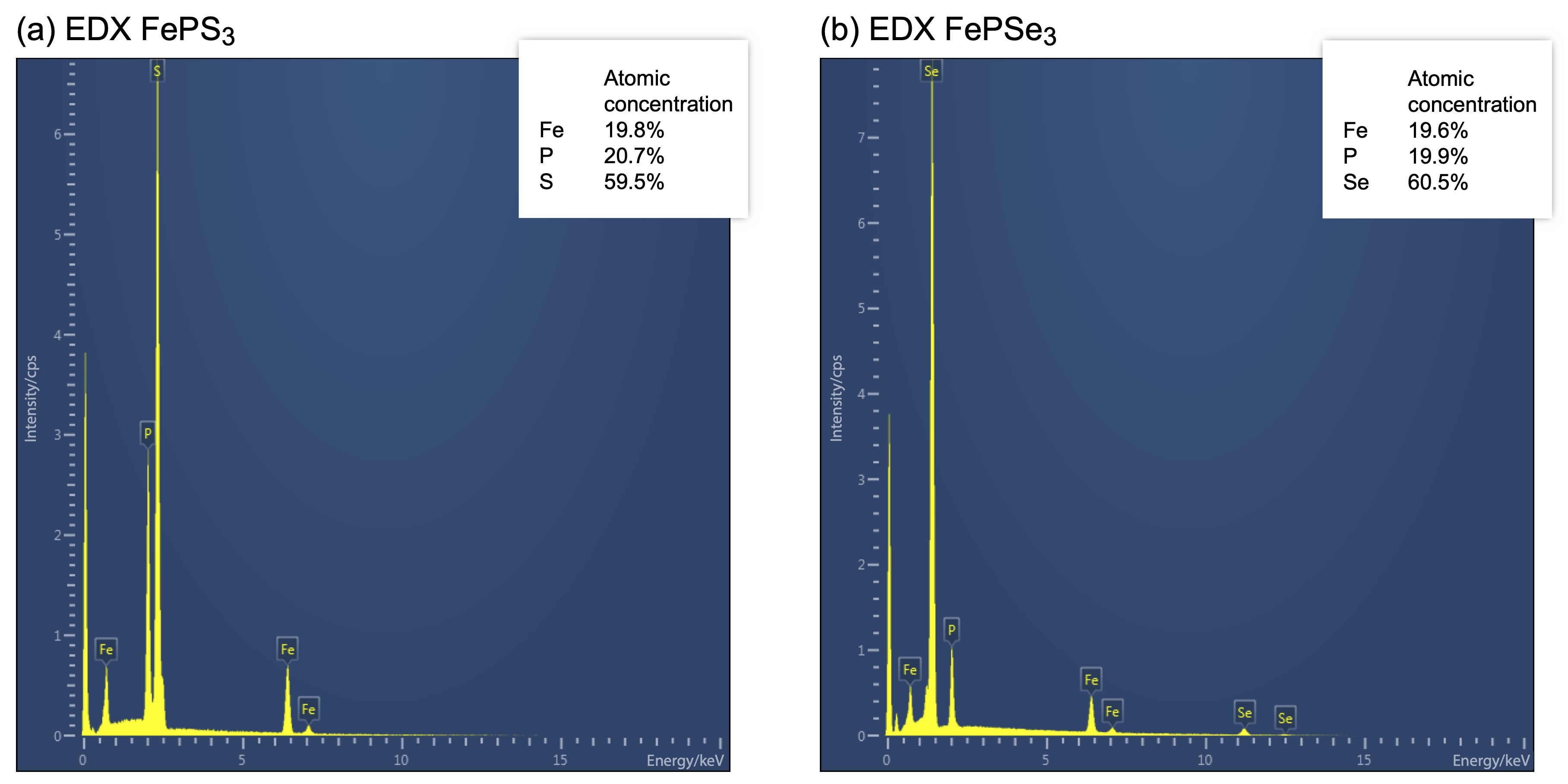}
\end{figure}
\noindent Fig.\,S3. Results of the EDX analysis for (a) FePS$_3$ and (b) FePSe$_3$.

\clearpage
\begin{figure}
\includegraphics[width=\textwidth]{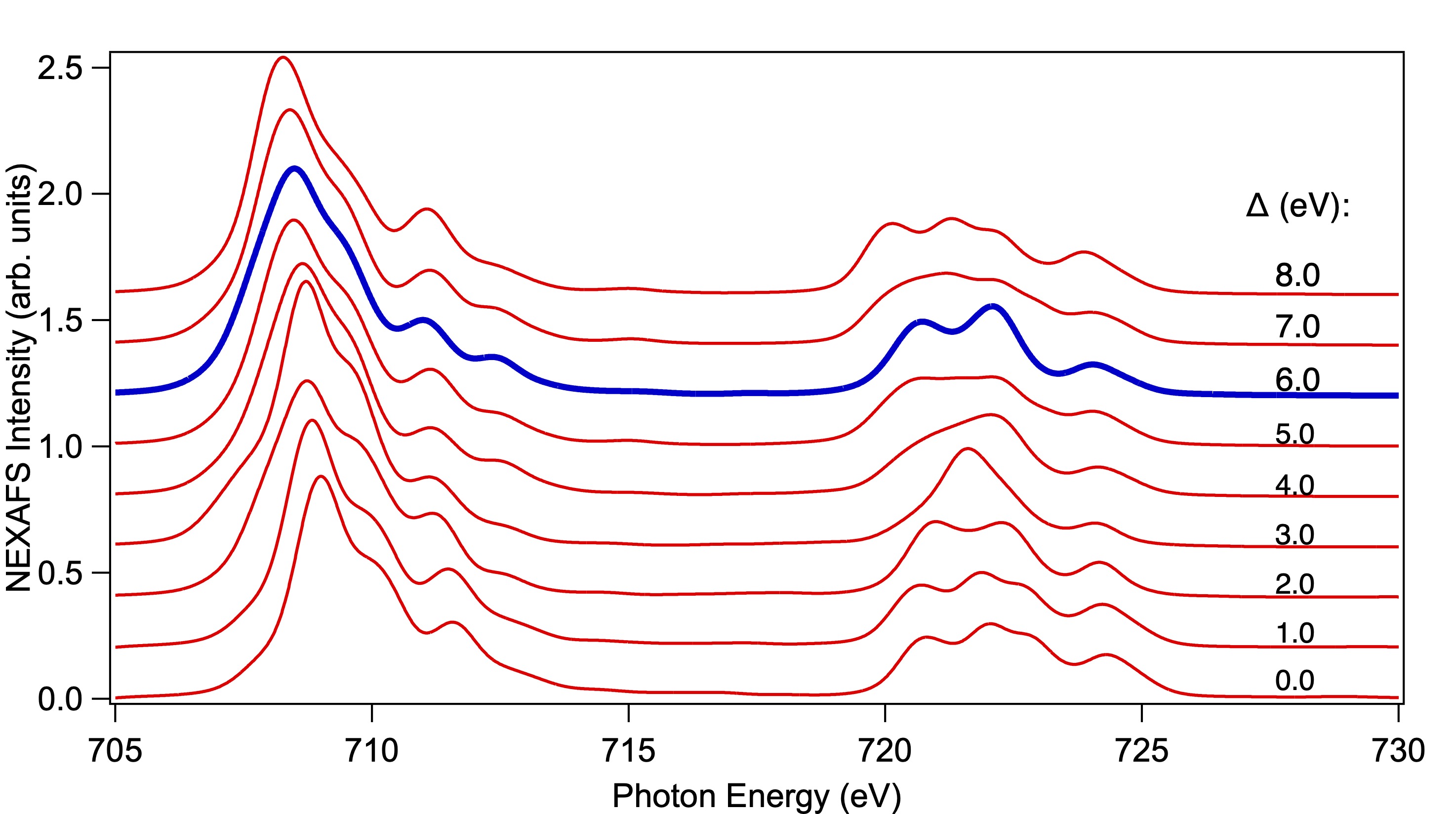}
\end{figure}
\noindent Fig.\,S4. Simulated Fe\,$L_{2,3}$ NEXAFS spectra of FePX$_3$ with $10Dq$$=$$0.3$\,eV, $U_{dd}$$=$$3$\,eV and different values of $\Delta$ marked for every spectra.

\end{document}